\documentclass[conference]{IEEEtran}
\IEEEoverridecommandlockouts
\usepackage{amsmath,amsfonts}
\usepackage{algorithm}
\usepackage{algpseudocode}
\usepackage{array}
\usepackage[caption=false,font=normalsize,labelfont=rm,textfont=rm]{subfig}
\usepackage{textcomp}
\usepackage{stfloats}
\usepackage{url}
\usepackage{verbatim}
\usepackage{graphicx}
\usepackage{cite}
\usepackage{amsmath}
\usepackage{amssymb}
\usepackage{amsfonts}
\usepackage{amsthm}                
\usepackage{booktabs}
\usepackage{caption} 
\usepackage{tabularray}
\usepackage{geometry}
\allowdisplaybreaks[4] 

\UseTblrLibrary{booktabs} % 配合 \toprule \midrule \bottomrule
\renewcommand{\baselinestretch}{0.99}
\begin{document}

\title{
Prospective Importance-based Cognitive Pilot Design for Semantic Sensing in ISAC Systems} 

\author{
    \IEEEauthorblockN{
        Guanping Shang$^*$, Anzheng Tang$^*$, Zeyan Zhuang$^*$, and Shenghui Song$^*$, \textit{Senior Member, IEEE}
    }
    \IEEEauthorblockA{
        $^*$Dept. of ECE, The Hong Kong University of Science and Technology, Hong Kong, China
    }
}
\maketitle

\begin{abstract}
Semantic sensing enables integrated sensing and communication (ISAC) systems to acquire environmental information relevant to downstream tasks. However, existing pilot design methods either rely on static pilot configurations or learn end-to-end pilot selection policies, without explicitly evaluating the semantic contribution of candidate pilot locations. In this work, we propose a cognitive semantic sensing framework for human activity recognition in orthogonal frequency division multiplexing (OFDM)-ISAC systems. 
Rather than directly learning a high-dimensional mapping from observations to pilot locations, we introduce a semantic importance weight (SIW) to measure the semantic contribution of each candidate pilot location. Leveraging the information bottleneck (IB) principle and the mutual information-minimum mean-square error (I-MMSE) relationship, SIW assigns high values to future time-frequency (TF) channel coefficients that are both difficult to infer from the preceding observations and valuable for downstream inference. To efficiently obtain the SIW, we further develop a conditional diffusion model that predicts the corresponding uncertainty through denoising covariance estimation. Experiments on a real-world Wi-Fi sensing dataset show that the proposed method consistently outperforms baselines across different signal-to-noise ratios (SNRs) and pilot budgets.
\end{abstract}


\section{Introduction}

Integrated sensing and communication (ISAC) enables wireless systems to support data transmission and environmental sensing using shared radio resources \cite{hkust2025review}. In conventional ISAC systems, sensing generally aims to recover physical information, such as channel states and target parameters \cite{physicalestimate, shang}. In contrast, semantic sensing focuses on extracting environmental information needed for downstream tasks, such as human activity recognition (HAR). 
This task-oriented objective makes pilot design particularly important because channel information acquired at different pilot locations contributes unequally to downstream inference. 
As a result, pilots should be allocated to acquire the most task-relevant channel information\cite{semisac}.

Existing semantic sensing methods typically apply a fixed pilot configuration learned from training data\cite{SemS}. Although such a configuration captures statistical regularities across the training distribution, it cannot adapt to specific channel variations. This limitation can be addressed by exploiting temporal dependence across sensing slots\cite{wang2020passive}. Preceding sparse observations can inform pilot allocation of pilots in subsequent slots, enabling pilot configurations to adapt to individual sensing instances.

Along this line of research, one approach is to directly learn the pilot configuration from preceding observations\cite{JPDCE}. However, learning a combinatorial allocation from high-dimensional channel observations is difficult and offers limited interpretability regarding individual TF channel coefficients\cite{ExplainableAI}. 
A more structured strategy is to first estimate the task importance of each candidate locations and then allocate pilots according to the estimated contributions. Similar importance-aware strategies have been adopted for resource allocation in task-oriented communication \cite{Importance-Aware1}. However, these resource allocation methods evaluate features that are already available before transmission, whereas pilot design requires assessing future TF channel coefficients before they are observed. The key challenge is therefore to estimate the prospective semantic contribution of each future TF element from preceding observations.

To address this challenge, we propose a cognitive pilot design framework for HAR task in orthogonal frequency division multiplexing (OFDM)-ISAC systems. Rather than directly predicting pilot locations, we introduce a semantic importance weight (SIW) that quantifies the prospective semantic contribution of each candidate pilot location. Based on the information bottleneck (IB) principle and the mutual information-minimum mean-square error (I-MMSE) relationship, the SIW accumulates the residual estimation uncertainty of each future TF channel coefficient according to its importance to the downstream task. It therefore assigns high values to channels that are both difficult to infer from preceding observations and valuable for semantic inference. 
To achieve efficient SIW estimation, we further develop a conditional diffusion model that predicts future TF channels and estimates their element-wise posterior uncertainty along its denoising trajectory. Pilots are then allocated to the candidate locations with the largest estimated SIWs. Experiments on the real-world CSI-Bench dataset\cite{dataset} demonstrate consistent improvements over uniform sampling, static semantic sensing, and direct cognitive pilot-learning baselines across different signal-to-noise ratios (SNRs) and pilot budgets.

\section{System Model and Problem Formulation}

\begin{figure}[tbp]
\centering 
\includegraphics[width=0.95\columnwidth]{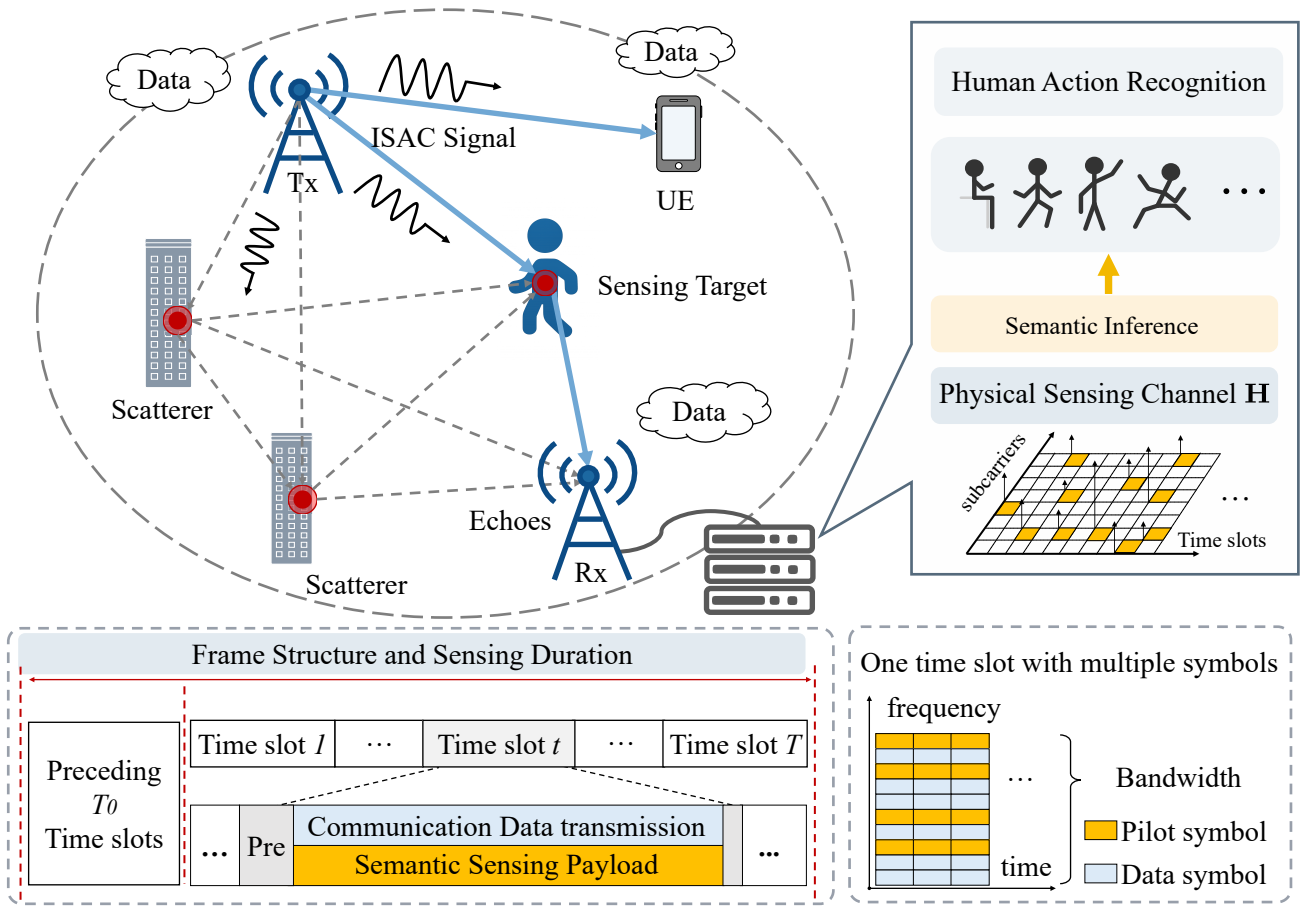}
\caption{Illustration of the OFDM-ISAC semantic sensing pilot and its frame architecture for HAR.}
\label{fig:HAR} 
\end{figure}

\subsection{OFDM-based Semantic Sensing}
\label{subsec:signal_model}
As illustrated in Fig.~\ref{fig:HAR}, we consider an OFDM-ISAC system that simultaneously supports communication and HAR sensing tasks. The ISAC frame spans $N_c$ subcarriers over $T$ time slots, forming an $N_c\times T$ TF grid. Out of the $N_c \times T$ TF elements, $N_p$ are assigned to sensing pilots, while the remaining are utilized for communication. 

Let $\mathbf X_{\rm loc}\in\{0,1\}^{N_c\times T}$ denote the binary pilot allocation matrix with $\|\mathbf X_{\rm loc}\|_0=N_p$, and let $\mathbf S_P\in\mathbb C^{N_c\times T}$ represent the unit-power ISAC symbols. Thus, the transmitted sensing pilot signal is expressed as
\begin{equation}
\mathbf X_P
=
\sqrt{E_P}\,
\mathbf X_{\rm loc}\odot\mathbf S_P
\in\mathbb C^{N_c\times T},
\label{eq:transmit_signal_x}
\end{equation}
where $\odot$ denotes element-wise multiplication and $E_P$ represents the per-pilot transmit power \cite{SemS,JPDCE}.

Let $\mathbf H\in\mathbb C^{N_c\times T}$ denote the effective sensing channel. Thus, the received pilot echoes on the TF grid are given by 
\begin{equation}
\mathbf Y_P
=
\mathbf H\odot\mathbf X_P+\mathbf V_P,
\label{eq:rx_signal}
\end{equation}
where $\mathbf V_P \in \mathbb{C}^{N_c \times T}$ denotes the effective sensing noise including additive white Gaussian noise (AWGN) and residual inter-carrier interference\cite{SemS}. 
Human motion induces structured channel variations\cite{wang2020passive}, which carry the underlying multipath delay and motion-induced Doppler characteristics for downstream HAR. 

The objective of semantic sensing is to extract the task-relevant information embedded in $\mathbf H$. Specifically, let $\mathcal S$ denote the target human-activity class associated with the sensing channel $\mathbf H$.  
Based on the observation $\mathbf Y_P$, the semantic receiver $\psi_{\mathbf\Theta}(\cdot)$ estimates $\mathcal S$ as \cite{SemS}  
\begin{equation}
\hat{\mathcal S}=\psi_{\mathbf\Theta}(\mathbf Y_P).
\end{equation}
The receiver side objective is to maximize the mutual information $I(\mathcal S;\hat{\mathcal S})$ between the actual and estimated semantic meaning, which unfortunately is difficult to evaluate for transmitter pilot design. We therefore introduce a tractable surrogate objective. Since the sensing process $\mathcal S\!\rightarrow\!\mathbf H\!\rightarrow\!\mathbf Y_P\!\rightarrow\!\hat{\mathcal S}$ forms a Markov chain, the data processing inequality gives \cite{DIP}
\begin{equation}
I(\mathcal S;\hat{\mathcal S})
\leq
I(\mathcal S;\mathbf Y_P).
\label{eq:dpi}
\end{equation} 
Thus, $I(\mathcal S;\mathbf Y_P)$ can be utilized as a surrogate objective for transmit-side pilot design.

\subsection{Cognitive Pilot Design Framework}
\label{subsec:framework}

% \footnote{More generally, such observations may also be obtained from other sensing modalities like visual sensors, while the pilot-based acquisition is adopted in our implementation according to the available dataset.}. 

% Conditioned on $\mathbf H_0$, the transmitter determines the transmitter determines the pilot 
% allocation matrix $\mathbf X_{\rm loc}$ for the subsequent $T$ sensing slots. 
In this paper, we propose to utilize the preceding observations to guide pilot design.
Specifically, for the first $T_0$ slots, a fixed pilot configuration is employed to obtain a sparse observation of the preceding channel $\mathcal H_0$.    
Following the feedback and pilot-configuration stage, the subsequent $T$ slots of cognitive sensing aims to select $N_p$ pilot locations that maximize the task-relevant mutual information $I(\mathcal{S}; \mathbf{Y_P})$. Specifically, given a realization $\mathbf H_0$ of $\mathcal H_0$, the cognitive pilot design is formulated as 
\begin{equation}
\begin{aligned}
&\mathbf X_{P}^{\star}(\mathbf H_0)
=
\arg\max_{\mathbf X_{\rm loc}}
I\!\left(
\mathcal S;\mathbf Y_P
\,\middle|\,
\mathcal H_0 =\mathbf H_0
\right)\\
\mathrm{s.t.}
&\quad
\mathbf X_{\rm loc}\in\{0,1\}^{N_c\times T},~
\|\mathbf X_{\rm loc}\|_0=N_p .
\end{aligned}
\label{eq:pointwise_opt}
\end{equation}
Directly solving \eqref{eq:pointwise_opt} requires determining a combinatorial pilot allocation from the high-dimensional observation, which is very challenging. To this end, we transform the problem into an element-wise contribution estimation problem in Sec.\ref{sec:Diffusion}.

\section{SIW-based Cognitive Pilot Design}\label{sec:Diffusion}

In this section, we develop the SIW-based cognitive pilot design scheme. For each candidate pilot location, the SIW jointly accounts for the residual uncertainty of the corresponding future TF channel coefficient and the benefit of accurately representing it for the downstream task. We estimate the former using a conditional diffusion network and covariance estimation, and approximate the latter through local classifier sensitivity under the IB formulation. The resulting SIWs are then used to allocate pilots to locations with highest prospective semantic contributions.

\subsection{Semantic Importance Weight Formulation}
\label{subsec:ib_siw}

To quantify the semantic contribution of individual channel coefficients, we introduce a Gaussian bottleneck representation of the future sensing channel\cite{DforA} 
\begin{equation}
\mathbf z_{\boldsymbol\Gamma}
=
\boldsymbol\Gamma^{1/2}\mathbf h+\mathbf n,
\qquad
\mathbf n\sim\mathcal{CN}(\mathbf 0,\mathbf I_D),
\label{eq:snr_bottleneck}
\end{equation}
where $\mathbf h=\operatorname{vec}(\mathbf H)\in\mathbb C^D$ with $D=N_c\times T$. 
The added Gaussian noise limits the information about $\mathbf h$ retained in $\mathbf z_{\boldsymbol{\Gamma}}$, while $\boldsymbol\Gamma=\operatorname{diag}(\gamma_1,\ldots,\gamma_D)\succeq0$ controls the amount of information retained about $\mathbf h$.  Specifically, a larger $\gamma_i$ produces a less noisy representation of $h_i$ and therefore preserves more of its information.

The amount of information retained about each channel coefficient should depend on its usefulness to the downstream task. 
We therefore determine $\boldsymbol{\Gamma}$ using the IB principle\cite{Dubois2020Learning}, which balances retaining information useful for inferring the semantic meaning $\mathcal{S}$ against limiting the amount of channel information retained:
\begin{equation}
\boldsymbol\Gamma^\star
=
\arg\min_{\boldsymbol\Gamma\succeq0}
\beta I(\mathbf z_{\boldsymbol\Gamma};\mathbf h \mid\mathcal H_0 =\mathbf H_0)-I(\mathbf z_{\boldsymbol\Gamma};\mathcal S \mid\mathcal H_0=\mathbf H_0). 
\label{eq:conditional_ib_lagrangian}
\end{equation}
The first term encourages $\mathbf z_{\boldsymbol\Gamma}$ to remain informative about the semantic variable $\mathcal{S}$, whereas the second penalizes the amount of information it retains about $\mathbf h$. The parameter $\beta>0$ controls this tradeoff. Consequently, a larger $\gamma_i^\star$ indicates that faithfully representing $ h_i$ provides more semantic benefit to justify the additional channel information retained.

To characterize how the retained channel information changes with the fidelity $\gamma_i$, we leverage the I-MMSE relationship \cite{I-MMSE}. It relates the derivative of $I(\mathbf h;\mathbf z_{\boldsymbol\Gamma}\mid\mathcal H_0 =\mathbf H_0)$  with respect to $\gamma_i$ to the conditional MMSE in estimating $\mathbf{h}_i$ from $\mathbf z_{\boldsymbol{\Gamma}}$ and $\mathbf H_0$. 
Specifically, for the Gaussian bottleneck representation $\mathbf z_{\boldsymbol\Gamma}$, the conditional I--MMSE identity \cite{I-MMSE} relates the mutual information to the corresponding MMSE as 
\begin{equation} 
\frac{\partial}
{\partial\gamma_i}
I(\mathbf h;\mathbf z_{\boldsymbol\Gamma}\mid\mathcal H_0 =\mathbf H_0)
=
\operatorname{mmse}_i(\boldsymbol\Gamma, \mathbf H_0),
\label{eq:coordinate_immse}
\end{equation}
where $
\operatorname{mmse}_i(\boldsymbol\Gamma,\mathbf H_0)
=
\mathbb E [|
h_i-\mathbb E[h_i\mid
\mathbf z_{\boldsymbol\Gamma},\mathcal H_0 =\mathbf H_0]|^2
\,|\,
\mathcal H_0 =\mathbf H_0]=
 \mathbb E_{\mathbf z_{\boldsymbol{\Gamma}}\mid\mathbf H_0}
 \left[
 [\boldsymbol\Sigma_{\boldsymbol{\Gamma}}
  (\mathbf z_{\boldsymbol{\Gamma}},\mathbf H_0)]_{i,i}
 \right]$. 

Evaluating \eqref{eq:coordinate_immse} over the $D$-dimensional space is computationally prohibitive. 
We therefore consider a common bottleneck $\boldsymbol\Gamma=\gamma\mathbf I_D$, where the marginal contribution of each channel element can be measured by element-wise MMSE 
\begin{equation}
m_i(\gamma,\mathbf H_0)
\triangleq
\operatorname{mmse}_i(\gamma\mathbf I_D,\mathbf H_0)
=
\mathbb E_{\mathbf z_\gamma\mid\mathbf H_0}
\left[
[\boldsymbol\Sigma_\gamma
(\mathbf z_\gamma,\mathbf H_0)]_{i,i}
\right].
\end{equation}
This common bottleneck formulation is characterized by the multi-level denoising trajectory introduced in Sec.\ref{subsec:conditional_diffusion_covariance}.

Correspondingly, based on the chain rule, the element-wise I--MMSE relation in \eqref{eq:coordinate_immse} satisfies 
\begin{align}
\frac{\mathrm d I(\mathbf h;\mathbf z_\gamma\mid\mathcal H_0 =\mathbf H_0)}
{\mathrm d\gamma}
&=
\sum_{i=1}^{D}
\frac{\partial I(\mathbf h;\mathbf z_{\boldsymbol\Gamma}\mid\mathcal H_0 =\mathbf H_0)}
{\partial\gamma_i}
\frac{\mathrm d\gamma_i}{\mathrm d\gamma}
\notag\\
&=
\sum_{i=1}^{D}m_i(\gamma,\mathbf H_0).
\label{eq:complex_scalar_immse}
\end{align}
Hence, $m_i(\gamma,\mathbf H_0)$ characterizes the contribution of the $i$-th channel coefficient to the increase in conditional channel information as $\gamma$ increases. The optimized bottleneck parameter $\gamma_i^\star$ determines the interval over which this contribution is accumulated under the IB tradeoff. We therefore define the SIW of the $i$-th pilot location as 
\begin{equation}
S_i(\mathbf H_0)
\triangleq
\int_0^{\gamma_i^\star}
m_i(\gamma,\mathbf H_0)\,\mathrm d\gamma. 
\label{eq:siwi_definition}
\end{equation}
Intuitively, the SIW can be interpreted as a semantic contribution: a large SIW identifies an element that remains difficult to predict from $\mathbf H_0$ and is relevant to the downstream task.

\subsection{Conditional Diffusion-based MMSE Estimation}
\label{subsec:conditional_diffusion_covariance}
Evaluating the SIW in \eqref{eq:siwi_definition} requires estimating the element-wise MMSE over different bottleneck levels, which is challenging for high-dimensional sensing channels. This challenge can be addressed by exploiting the diffusion trajectory, which provides a sequence of noisy observations at different effective SNR levels. We therefore develop a diffusion model conditioned on $\mathbf H_0$ to generate a multi-level denoising trajectory from which the posterior covariance is estimated.

\subsubsection{Conditional MMSE Estimation}
We consider a normalized diffusion process 
$
\mathbf z_\tau =
\mathbf h+\sigma_\tau\boldsymbol\epsilon_\tau,
$
where $\mathbf z_\tau$ denotes the noisy observation at diffusion timestep $\tau$ with the SNR $\gamma_\tau=\sigma_\tau^{-2}$, and $\boldsymbol\epsilon_\tau\sim\mathcal{CN}(\mathbf0,\mathbf I_D)$\cite{DDPM}. 
After normalization by $\sigma_\tau$, this observation is equivalent to the Gaussian bottleneck in \eqref{eq:snr_bottleneck} with $\gamma=\gamma_\tau$.

According to the second-order Tweedie identity, the posterior covariance of $\mathbf h$ is implicitly reflected in the denoising process\cite{Tweedie,FreeHunch}. 
Specifically, let $\mathbf z$ denote a realization of
$\mathbf z_\tau$ and $\boldsymbol\mu_\tau(\mathbf z,\mathbf H_0)
=\mathbb E[\mathbf h|\mathbf z_\tau=\mathbf z,\mathcal{H}_0 =\mathbf H_0]$
represent the conditional denoising mean. The corresponding posterior covariance satisfies
\begin{equation}
\boldsymbol\Sigma_\tau(\mathbf z,\mathbf H_0)
=
\sigma_\tau^2
\frac{\partial
\boldsymbol\mu_\tau(\mathbf z,\mathbf H_0)}
{\partial \mathbf z},
\label{eq:tweedie_covariance_jacobian}
\end{equation}
where the derivative denotes the Jacobian of the conditional denoising mean with respect to the noisy observation.

Directly evaluating the Jacobian, however, is computationally expensive. 
We therefore adopt Free Hunch (FH) \cite{FreeHunch}, which recursively estimates the posterior covariance $\widehat{\boldsymbol\Sigma}^{\rm FH}_\tau$ from consecutive states along the reverse denoising trajectory. The element-wise MMSE is consequently estimated as $
\widehat m_i(\gamma_\tau,\mathbf H_0)
=
\left[
\widehat{\boldsymbol\Sigma}^{\rm FH}_\tau
\right]_{i,i}.
$ 
Therefore, the diffusion--FH procedure provides the element-wise conditional MMSE required to evaluate \eqref{eq:siwi_definition}, without explicitly constructing the denoiser Jacobian.

\subsubsection{Conditional Diffusion Model}
\label{subsubsec:conditional_diffusion_network}

\begin{figure}[tbp]
\centering
\includegraphics[width=2.8in]{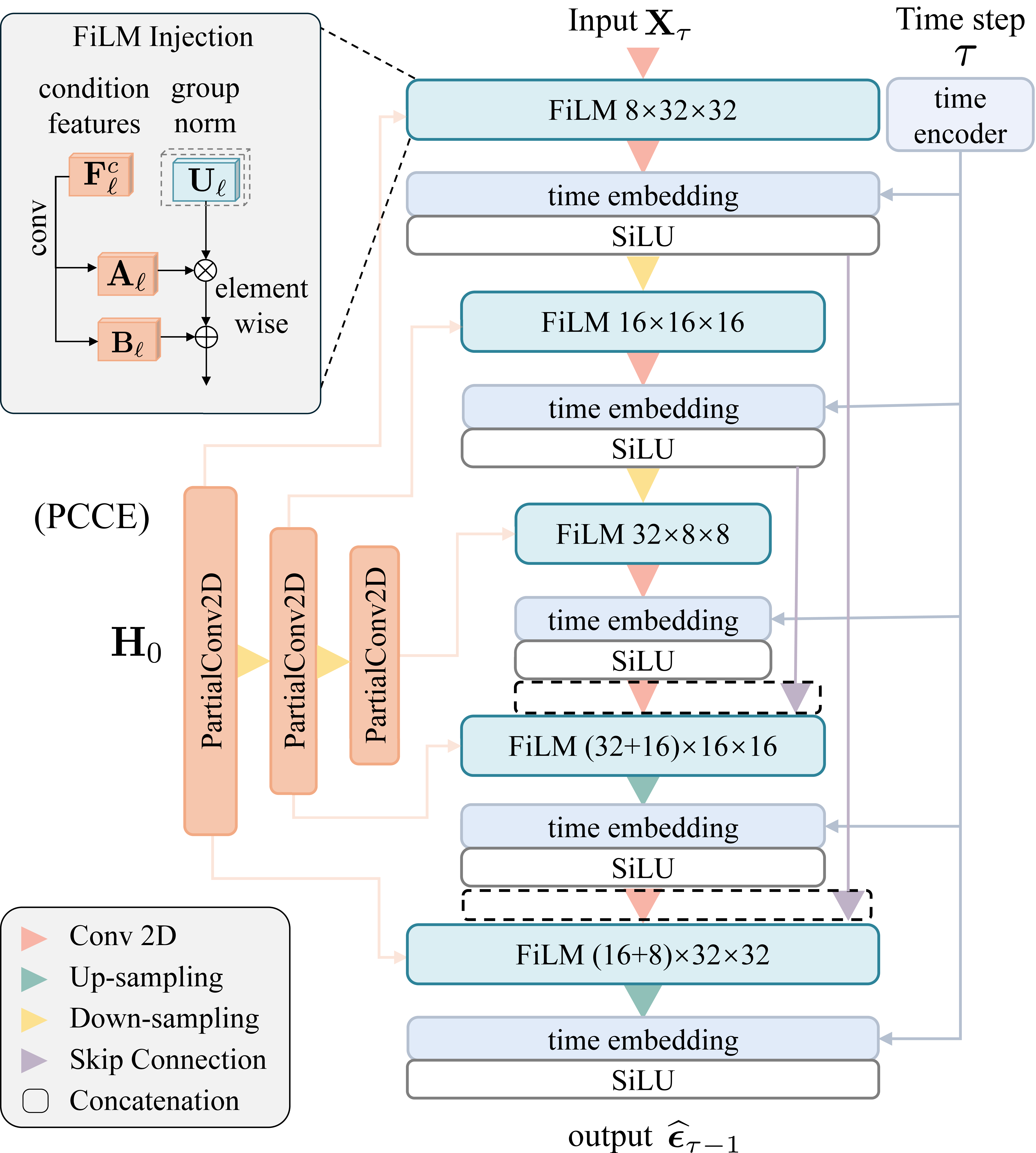}
\caption{Illustration of the conditional
diffusion model.}
\label{fig:conditional_diffusion_architecture}
\end{figure}

To implement the conditional denoising process required above, we develop a conditional diffusion network shown in Fig.~\ref{fig:conditional_diffusion_architecture}. The network takes $\mathbf H_0$ as conditional input and performs the reverse diffusion process. 

For the HAR task considered in this work, we use the channel magnitude $|\mathbf H|$ as the sensing input, where $|\cdot|$ denotes the element-wise absolute value. Accordingly, the conditional diffusion model operates on the real-valued amplitude representation\footnote{
The adopted CSI-Bench dataset provides amplitude-only CSI for HAR~\cite{dataset}. The amplitude variation captures action-dependent multipath and frequency-selective variations that support HAR task, whereas the measured phase can be sensitive to oscillator drifts and synchronization offsets across devices. Moreover, the proposed method also supports complex-valued channel cases via two real-valued input channels.}.  
Specifically, let $|\mathbf H_0|\in\mathbb R^{N_c\times T_0}$ denote the preceding observation. 
As illustrated in Fig.~\ref{fig:conditional_diffusion_architecture}, the diffusion model consists of a partial convolution conditional encoder (PCCE) and a U-Net main branch. 
The PCCE excludes zero entries of sparse pilots from each receptive field. Specifically, let
$(\mathbf F_{\ell-1}^{c},\mathbf M_{\ell-1})$ denote the
condition feature and its validity mask at layer $\ell-1$. The PCCE operation is expressed as
\begin{align}
  \mathbf F_\ell^{c}&=\left[\mathbf W_\ell^{c}\circledast \left(
 \mathbf F_{\ell-1}^{c}\odot\mathbf M_{\ell-1}\right)
 \right]\odot\frac{K_\ell}{\max(\mathbf Q_\ell, \mathbf{1}_{})}+\mathbf b_\ell^{c} \odot\mathbf M_\ell, \nonumber\\\mathbf Q_\ell&=\mathbf 1_\ell\circledast \mathbf M_{\ell-1},\qquad\mathbf M_\ell=\mathbb I(\mathbf Q_\ell>0),
 \label{eq:partial_convolution}
\end{align}
where $\circledast$ denotes convolution operation, $\mathbf W_\ell^c$ and $\mathbf b_\ell^c$ denote the convolution kernel and bias, respectively, $\mathbf 1_\ell$ denotes an all-one kernel, $K_\ell$ denotes the number of elements in the receptive field, and $\mathbb I(\cdot)$ denotes the indicator function.  
% Starting from
% $(\mathbf F_0^{c},\mathbf M_0)=(|\mathbf H_0|,\mathbf M_0)$, where $\mathbf M_0\in\{0,1\}^{N_c\times T_0}$ identifies the non-zero entries of the preceding observation, successive partial convolution and downsampling produce multi-scale condition features $\{\mathbf F_\ell^{c}\}_{\ell=1}^{L}$ aligned with the $L$ layers main branch.
The main branch adopts an encoder--decoder U-Net with downsampling and upsampling, where the condition feature is injected into
the corresponding U-Net feature through spatial feature-wise
linear modulation (FiLM) at each layer \cite{film}. 
% During the training process, a diffusion state is generated from the original channel $\mathbf X=\mathcal{T}(\mathbf H)=|\mathbf H| \in \mathbb{R}^{N_c \times T}$ according to the standard forward
% process. 

Let $\mathbf X_0=|\mathbf H|$ denote the clean channel amplitude, and $\mathbf X_\tau$ represent its noisy version at diffusion timestep $\tau$. 
The conditional diffusion network 
$\boldsymbol\epsilon_{\boldsymbol\theta_d}$ 
predicts the Gaussian noise $\boldsymbol\epsilon_\tau$ with parameters $\boldsymbol\theta_d$ trained by 
\begin{equation}
\boldsymbol\theta_d^\star
=
\arg\min_{\boldsymbol\theta_d}
\mathbb E_{\tau,\boldsymbol\epsilon_\tau}
\left[
\left\|
\boldsymbol\epsilon_\tau-
\boldsymbol\epsilon_{\boldsymbol\theta_d}
(\mathbf X_\tau,\tau,|\mathbf H_0|)
\right\|_2^2
\right].
\label{eq:conditional_diffusion_loss}
\end{equation}

During inference, the denoised estimates produced at consecutive reverse steps are supplied to FH. Hence, the diffusion model
and FH jointly predict the posterior covariance from a single
reverse-diffusion trajectory. 

\subsection{Task-Dependent Bottleneck Approximation}
\label{subsec:gradient_gamma}
Besides the conditional MMSE estimated in Sec.\ref{subsec:conditional_diffusion_covariance}, evaluating the SIW also requires the element-wise task relevance $\gamma_i^\star$ defined by \eqref{eq:conditional_ib_lagrangian}. Since the mutual-information terms in \eqref{eq:conditional_ib_lagrangian} are difficult to evaluate directly, we adopt a tractable surrogate. Specifically, the negative sufficiency term $-I(\mathcal S;\mathbf z_{\boldsymbol\Gamma}
\mid\mathbf H_0)$ is approximated by the expected HAR cross-entropy (CE) loss $\mathcal L_{\mathrm{task}}$, while the minimality term $I(\mathbf h;\mathbf z_{\boldsymbol\Gamma} \mid\mathbf H_0)$ is represented by an element-wise logarithmic regularizer on $\gamma_i$ \cite{DforA}. The resulting surrogate optimization is given by  
\begin{equation}
 \widehat{\boldsymbol\Gamma} 
 =
 \arg\min_{\boldsymbol\Gamma\succ\mathbf0}
 \mathbb E_{\boldsymbol\varepsilon}
 \left[
 \mathcal L_{\mathrm{task}}
 \left(
\mathbf z_{\boldsymbol\Gamma}
 \right)
 \right]
 +
 \beta\sum_{i=1}^{D}\log\gamma_i .
 \label{eq:empirical_ib_gamma}
\end{equation} 
To avoid iterative optimization of \eqref{eq:empirical_ib_gamma} for each sensing instance, we derive a closed-form element-wise approximation. Let $\mathbf x=\operatorname{vec}(|\mathbf H|)$ denote the vectorized channel amplitude. Using the local perturbation analysis in Appendix A, the bottleneck is approximated as 
\begin{equation}
\widehat\gamma_i^\star
=
\operatorname{clip}
\left(
{g_i^2}/{2\beta},
\gamma_{\min},
\gamma_{\max}
\right),
\label{eq:gradient_gamma_star}
\end{equation}
where
$g_i=\left.\partial\mathcal L_{\rm task}(\mathbf x)/\partial x_i\right|_{\mathbf x=\widehat{\mathbf x}_0}$
is evaluated using the classifier $\psi_\rho $ with parameters $\rho$ pretrained on fully sampled channels. 
The classification result based on the diffusion estimate $\widehat{\mathbf x}_0$ serves as the pseudo-label in $\mathcal L_{\rm task}$, requiring neither the true future channel nor its ground-truth label.
With the task-dependent bottleneck level $\widehat{\gamma}_i^\star$ obtained above and the conditional MMSE $\widehat m_i(\gamma_\tau,\mathbf H_0)$ estimated in Sec.\ref{subsec:conditional_diffusion_covariance}, all quantities required to evaluate the SIW in (11) are now available. We next combine these two components to estimate the SIWs and translate them into a discrete pilot allocation.

\subsection{SIW Estimation and Pilot Allocation}

\begin{figure}[tbp]
\centering
\includegraphics[width=\columnwidth,height=1in]{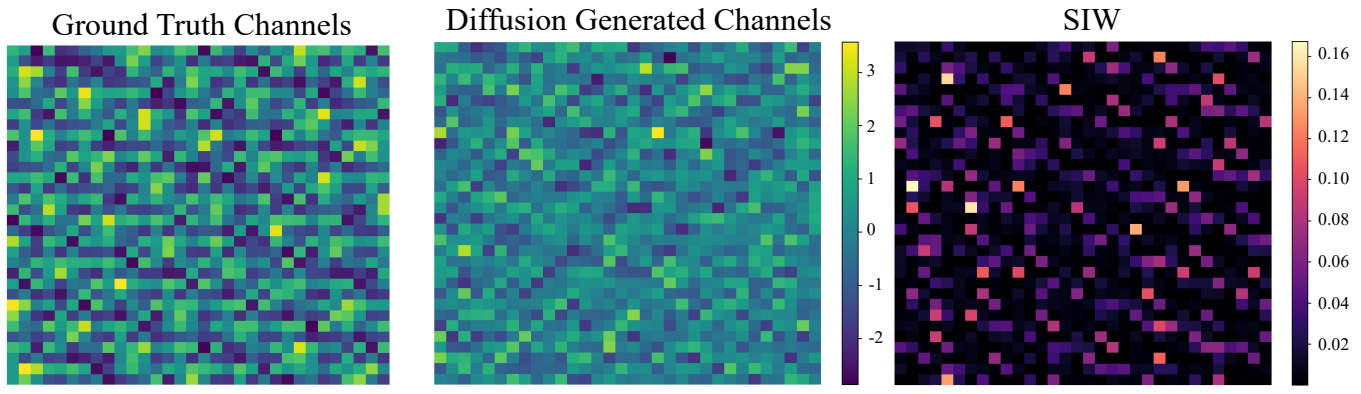}
\caption{Visualization of the CSI and SIW results.}
\label{fig:siw_visualization} 
\end{figure}

With the element-wise MMSE $\widehat m_i(\gamma_\tau,\mathbf H_0)$ and task relevance $\widehat{\gamma}_i^\star$ estimated in Sec.\ref{subsec:conditional_diffusion_covariance} and \ref{subsec:gradient_gamma}, the SIW estimation $\widehat S_i(\mathbf H_0)$ is obtained via numerical integrating \eqref{eq:siwi_definition} over the discrete $\gamma_\tau$, yielding an estimated prospective semantic contribution for each candidate pilot location.
Fig.~\ref{fig:siw_visualization} visualizes the ground-truth channels, the diffusion generated channels, and the estimated SIW. 
The diffusion prediction preserves the major structures of the future channel. 
The estimated SIW exhibits a highly non-uniform distribution, highlighting TF locations with larger estimated SIWs.

Pilots are allocated to the $N_p$ TF grids with the largest SIWs. 
Since $m_i(\gamma,\mathbf H_0)$ is obtained from the joint posterior covariance, statistical dependencies among the TF channel elements are implicitly reflected in the resulting SIWs. 
Accordingly, the pilot configuration is given by 
\begin{equation}
\left[\operatorname{vec}\!\left(\mathbf X_{\rm loc}\right)\right]_i
=
\mathbb I\!\left(
\widehat S_i(\mathbf H_0)
\in
\operatorname{Top}_{N_p}
\left\{\widehat S_j(\mathbf H_0)\right\}_{j=1}^{D}
\right), 
\label{eq:topk_pilot_selection}
\end{equation}

\section{Experimental Results}
\subsection{Experimental Setup}
We evaluate the proposed framework on in-the-wild Wi-Fi sensing dataset CSI-Bench \cite{dataset}, which contains five HAR classes, following the official CSI-Bench training, validation, and test split with a ratio of 70/15/15. 
Each TF frame comprises $32\times32$ TF grids, with each time slot lasting 156.3~ms and the frequency bandwidth spanning 546.8~kHz. The sparse channel in the first $T_0=3$ slots serves as the preceding sparse observation, one slot is reserved for feedback and pilot configuration, and the remaining $T=28$ slots form the future sensing channels, yielding $D=896$. 
The conditional diffusion model uses a three-scale PCCE/U-Net with feature dimensions $8$, $16$, and $32$, and is trained for 50 epochs with Adam at an initial learning rate of $10^{-4}$. 
We employed 40 diffusion timesteps spaced from $10^{-4}$ to $30$ and 20 FH covariance estimation steps for SIW integration.

We compare the proposed framework with the following baselines. 

\textbf{Full Sampling:} This baseline places a pilot at every TF location, providing complete channel observations and serving as an upper bound for pilot designs.

\textbf{Uniform Sampling:} This baseline uniformly distributes $N_p$ pilots over the TF grid.

\textbf{Semantic Sensing (SemS)~\cite{SemS}:} 
SemS is a feedback-free baseline using a trainable Gumbel--Softmax estimator, which learns a static, input-independent pilot configuration. 

\textbf{JPDCE~\cite{JPDCE}:}
JPDCE is an input-dependent method that directly maps the observations to pilot location using a deep network, thereby realizing an instance-adaptive pilot configuration via end-to-end training. 

\textbf{Single-factor SIW ablations:}
To examine the effect of MMSE and IB in SIW,
we consider two ablation variants. SIW w/o IB selects pilot locations using only
the marginal posterior MMSE $\widehat m_i(\gamma_\tau,\mathbf H_0)$,
without IB-based task relevance $\widehat \gamma_i$. Conversely, SIW w/o MMSE uses only
the $\widehat \gamma_i$, without considering MMSE.

For a fair comparison, all methods use the same preceding observations $\mathbf H_0$, unit-power pilot symbols, pretrained MLP receiver $\psi_\rho(\cdot)$, and LS channel estimation\cite{SemS}.

\begin{figure}[t]
\centering
\subfloat[Accuracy]{%
  \includegraphics[width=0.24\textwidth]{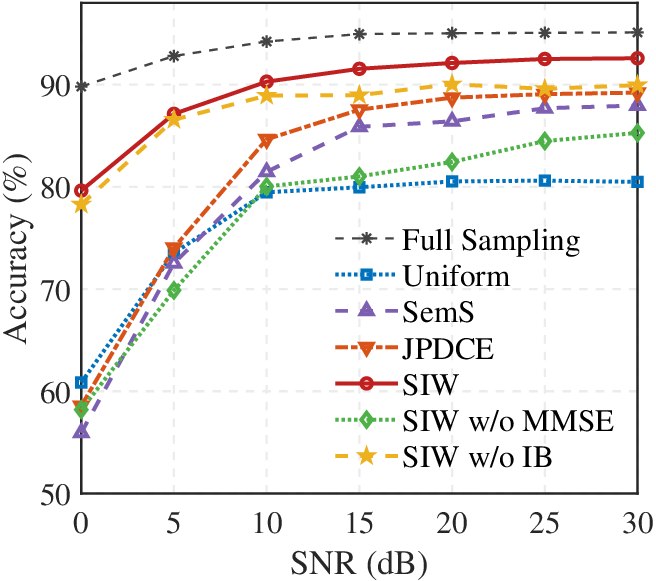}%
}
\subfloat[F1 score]{%
  \includegraphics[width=0.24\textwidth]{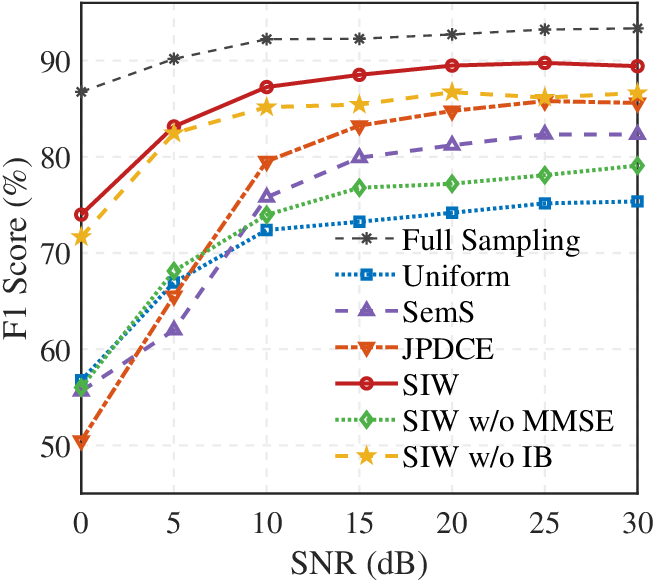}%
}
\caption{Accuracy and F1 Score versus SNR.}
\label{fig:SNR}
\end{figure}

\subsection{Semantic Sensing Performance}
Fig.~\ref{fig:SNR} reports the sensing accuracy and F1 score versus
the average SNR with $N_p=256$. The accuracy of all sensing methods
improves with SNR increasing. The SIW method consistently provides the best performance over the entire SNR range, and the advantage is especially
pronounced in the low SNR regime. 
SIW w/o MMSE performs relatively poorly because it cannot distinguish between task-sensitive coefficients that remain uncertain and those that are predictable from the preceding observations. 
Although SIW w/o IB is not explicitly task-dependent, it prioritizes CSI fidelity at the physical level through MMSE estimation, thereby achieving competitive performance. These results demonstrate that SIW provides informative guidance for pilot allocation.

\subsection{Communication-Sensing Trade-off}
\begin{figure}[t]
\centering
\subfloat[Accuracy]{%
  \includegraphics[width=0.24\textwidth]{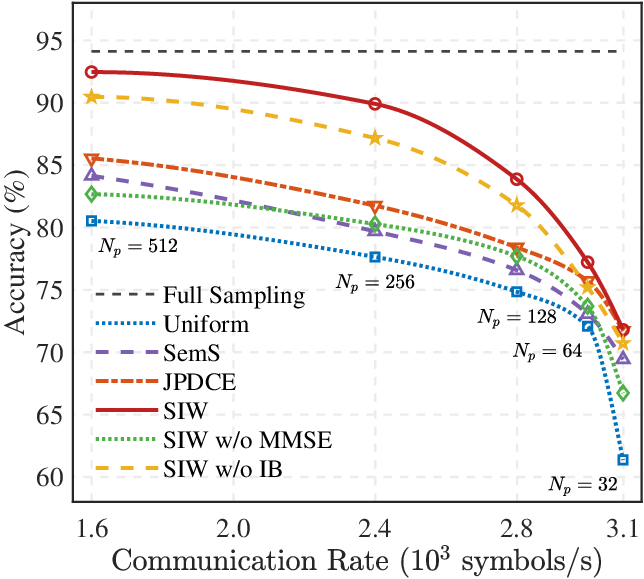}%
}
\subfloat[F1 score]{%
  \includegraphics[width=0.24\textwidth]{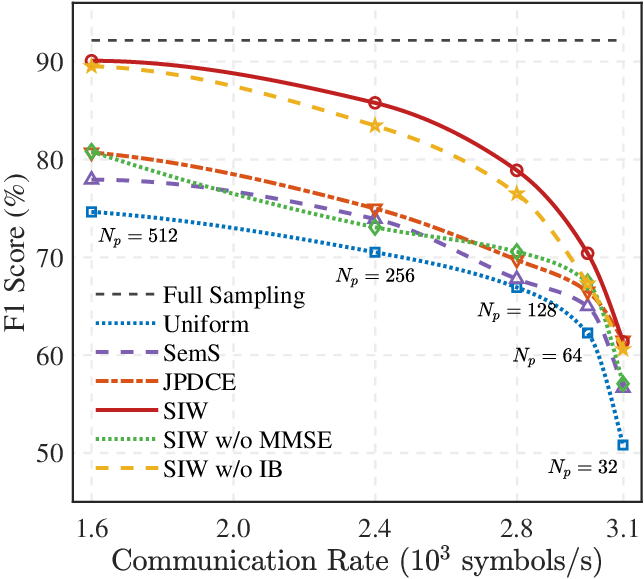}%
}
\caption{Accuracy and F1 score versus achievable communication symbol rate at different sensing pilot budgets.}
\label{fig:Np} 
\end{figure}

Fig.~\ref{fig:Np} evaluates the trade-off between semantic sensing performance and the available communication symbol rate versus the pilot budget $N_p$ at SNR $=20$ dB. A higher communication rate leaves fewer TF resources for sensing pilots, resulting in degraded HAR performance for all methods. Conversely, reducing the communication rate allows more pilots to be allocated for sensing and improves both accuracy and F1 score. Across the entire operating range, the proposed SIW-based design consistently achieves the best sensing performance among the sparse-pilot schemes and approaches the Full Sampling bound as more sensing resources become available. These results demonstrate that SIW enables a more effective allocation of the shared TF resources, providing improved semantic sensing performance for a given resource budget.

\subsection{Pilot Allocation Result Analysis}
\begin{figure}[t]
\centering
\includegraphics[width=\columnwidth]{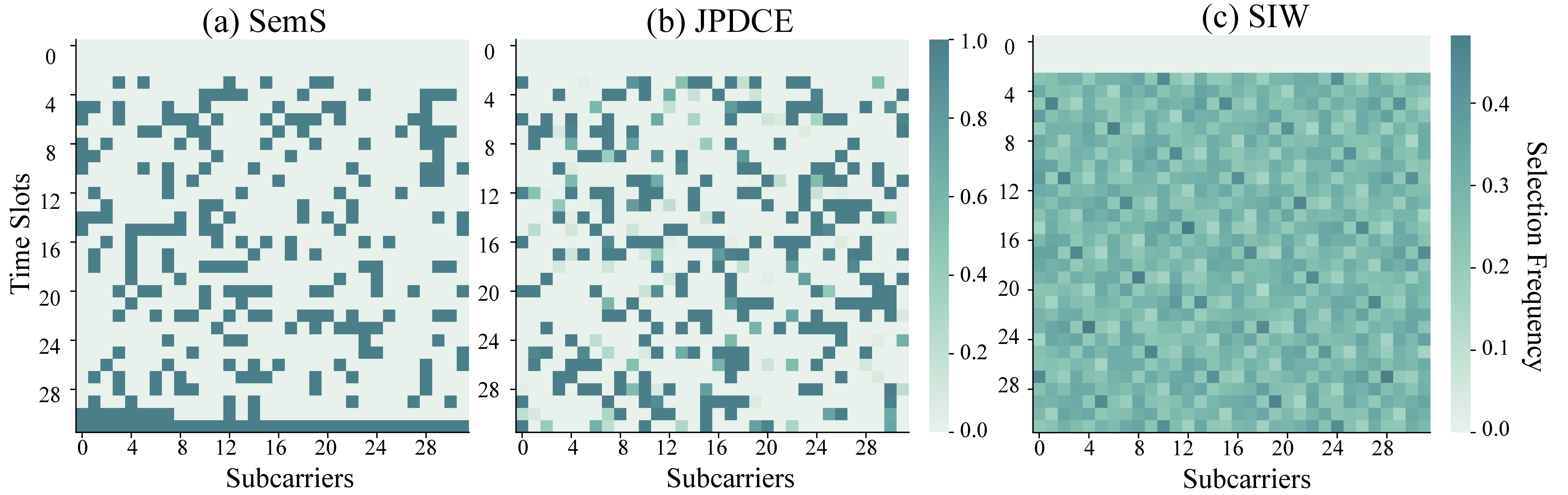}
\caption{Pilot selection frequencies of different methods over the test set.}
\label{fig:heatmap_all} 
\end{figure}

Fig.~\ref{fig:heatmap_all} further compares pilot selection frequency across different methods. Each grid shows the selection frequency of the corresponding TF location over the test set. Since SemS employs a fixed pilot pattern across all sensing instances, its pilot-selection frequency at each TF location is either zero or one.
Although JPDCE generates an input-dependent pilot configuration, its selections remain concentrated on a limited set of TF locations across sensing instances, indicating limited variation in its pilot allocations.
This behavior suggests that the End-to-End learning method tends to favor TF locations that provide good average performance over the training distribution, leading to limited instance-specific adaptation. 
In contrast, the proposed SIW method exhibits a more broadly distributed selection frequency map, which captures the instance-dependent channel variations and enables stronger instance-adaptive pilot configuration.

\subsection{Semantic Representation Visualization} 
\begin{figure}[t]
\centering
\subfloat[Full Sampling]{%
  \includegraphics[width=0.47\columnwidth]{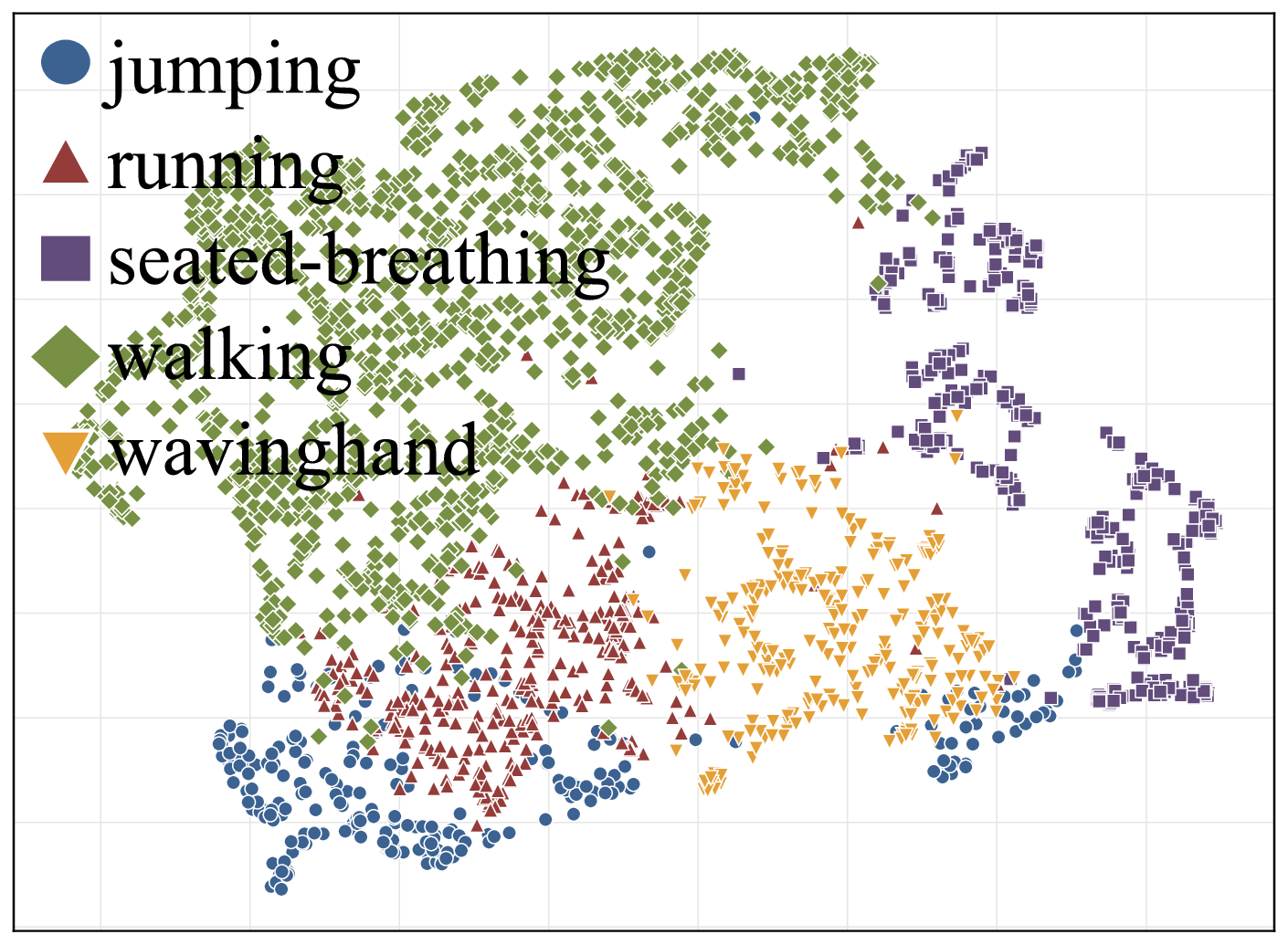}%
}\hspace{1mm}% 控制左右紧密程度
\subfloat[Proposed]{%
  \includegraphics[width=0.47\columnwidth]{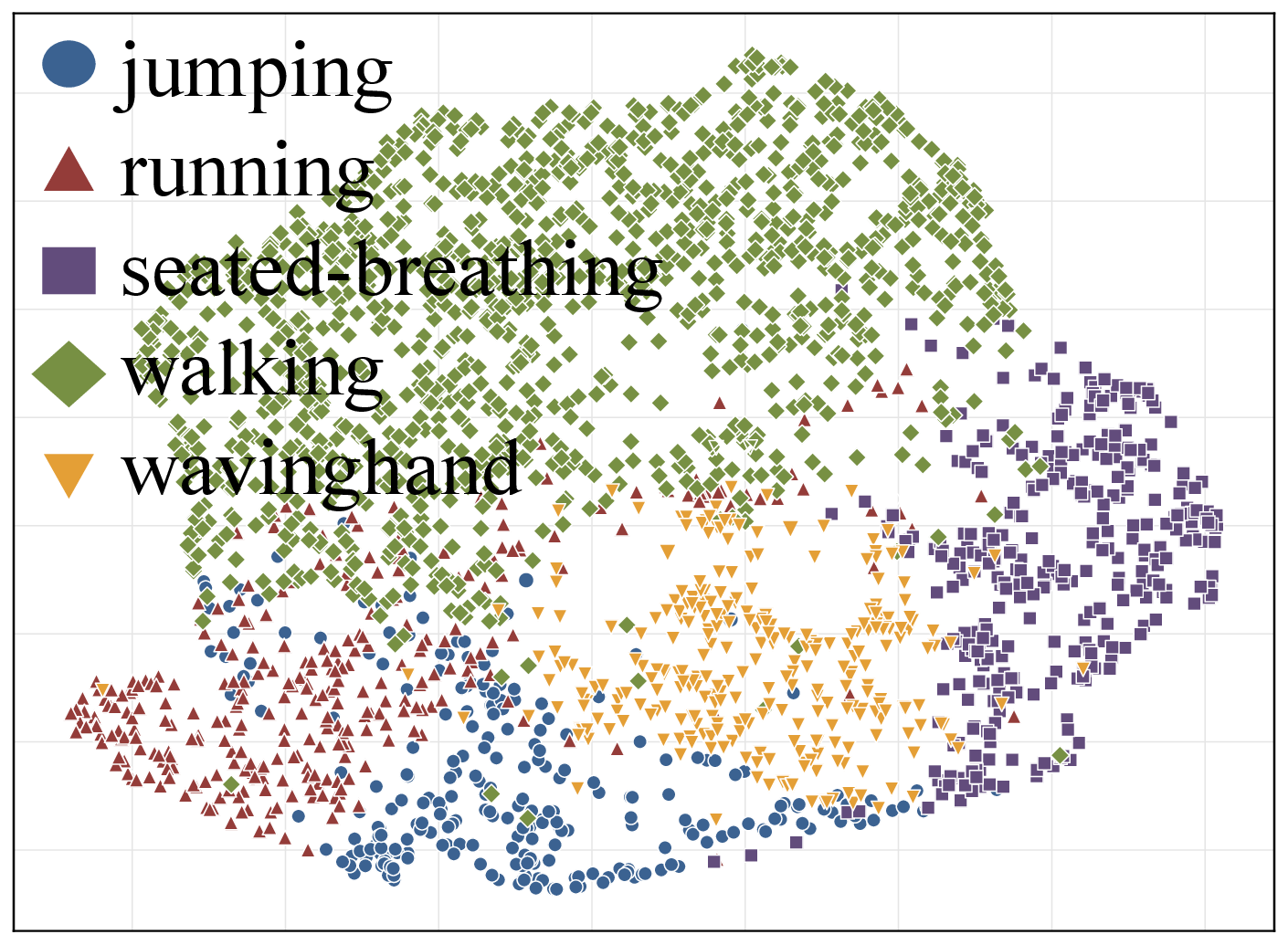}%
}\\[-2pt] % 负间距使上下更紧凑
\subfloat[JPDCE]{%
  \includegraphics[width=0.47\columnwidth]{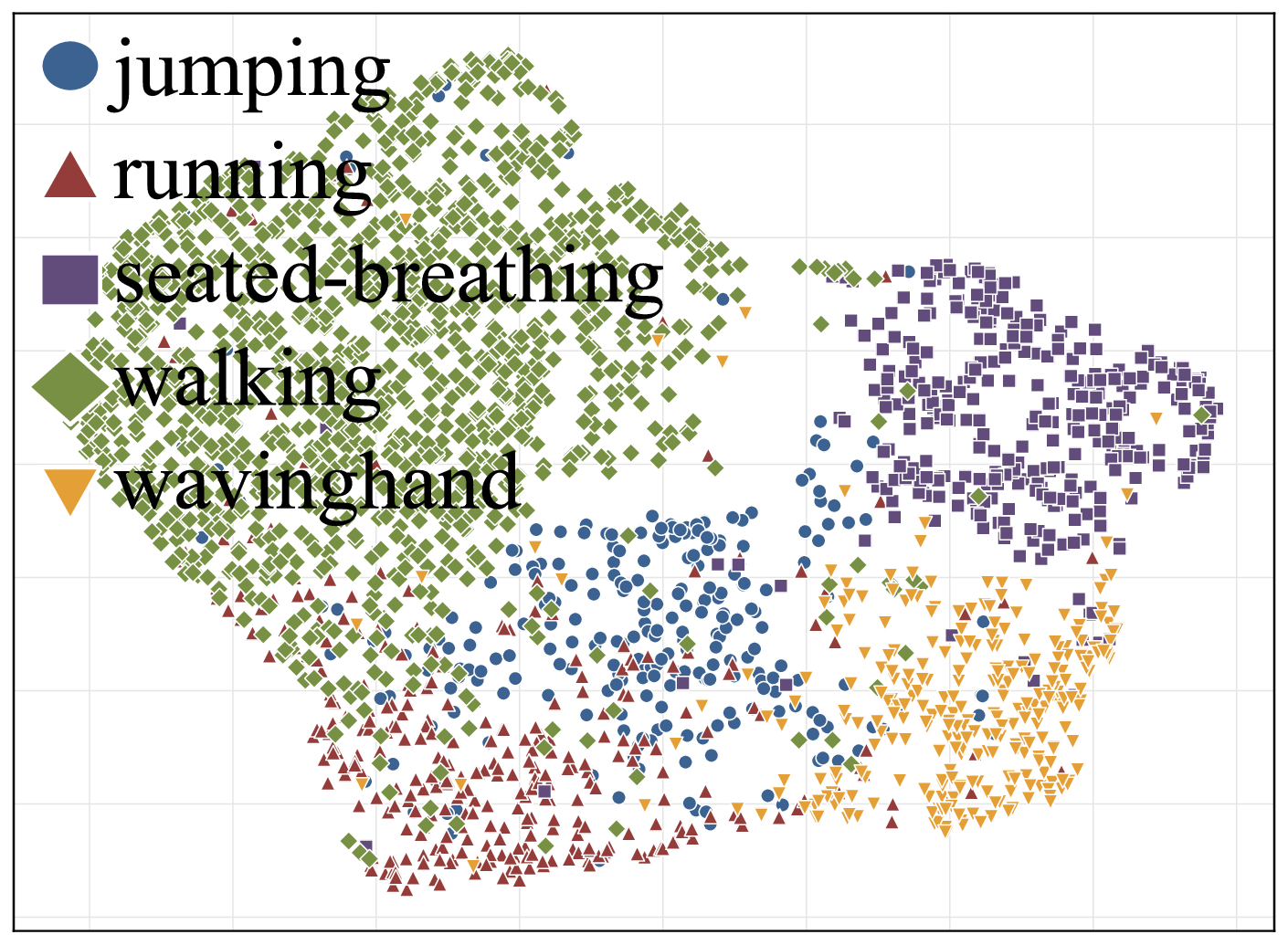}%
}\hspace{1mm}%
\subfloat[SemS]{%
  \includegraphics[width=0.47\columnwidth]{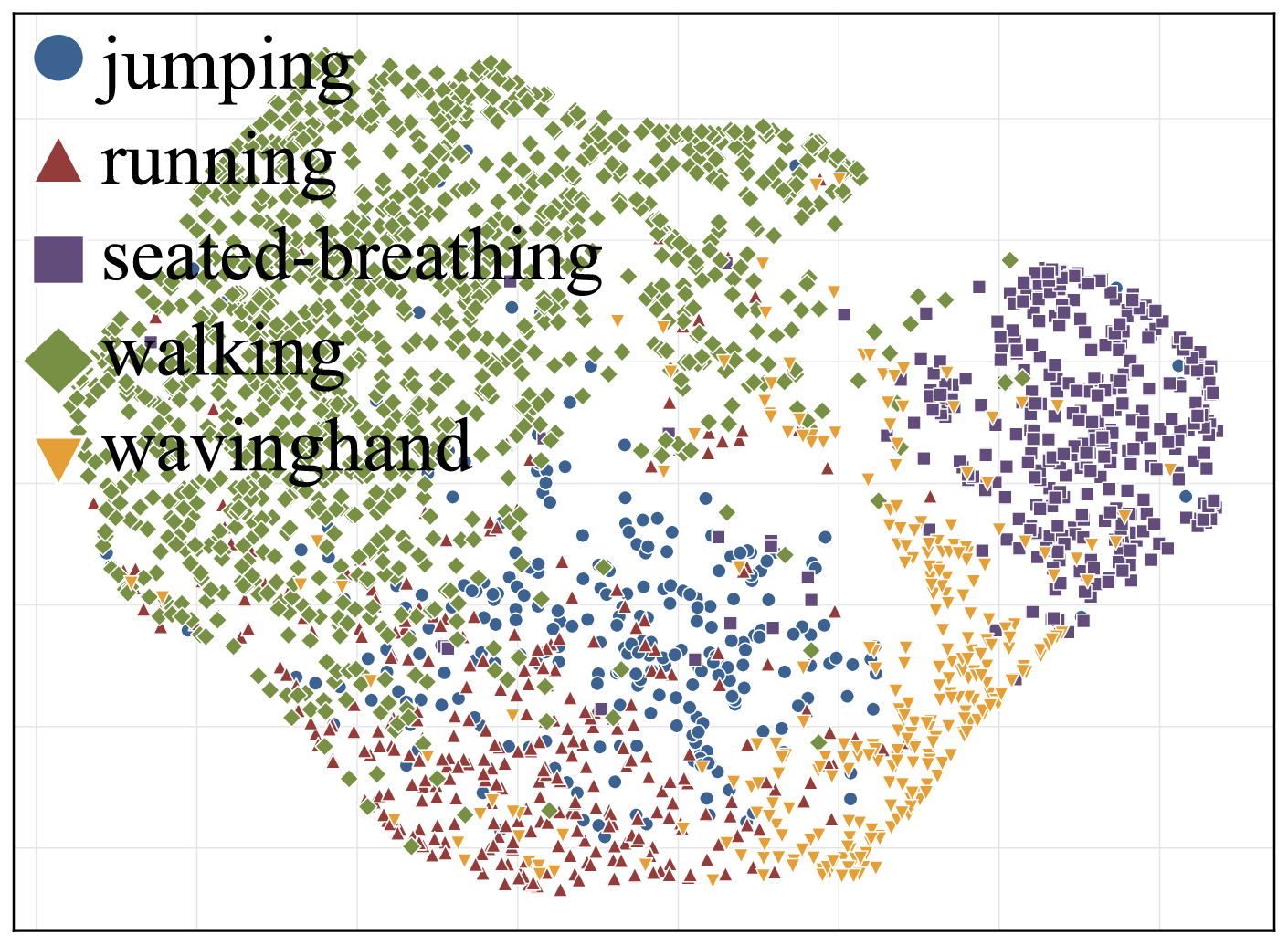}%
}
\caption{UMAP visualization of the semantic sensing features with
different pilot design methods.}
\label{fig:umap_all}
\end{figure}

To visualize the learned semantic representation,
Fig.~\ref{fig:umap_all} projects the classifier features into two
dimensions by Uniform Manifold Approximation and Projection (UMAP). 
The Full Sampling baseline yields the best distributions. The proposed method most closely preserves such action-dependent structure, where the intra-class compactness and inter-class separation in the projected feature space are well-maintained. In comparison, JPDCE and SemS exhibit noticeable overlap among several classes, indicating that part of the discriminative channel information is lost under their pilot configurations. 
% Uniform sampling produces the most scattered and entangled representation, with substantial inter-class mixing and weakened class-specific structure. 

\subsection{Computational Complexity}
\begin{table}[h]
\centering
\caption{Model Complexity Analysis}
\label{tab:model_complexity_tblr}
\scriptsize
\begin{tblr}{
    width = \linewidth,
    colspec = {l X[c] X[c] X[c] X[c] X[c]}, % 第一列左对齐，其余列自适应居中
    row{1} = {font=\bfseries},              % 第一行自动加粗
}
\toprule
% 如果不需要第一列的分组名称，可以把下面带 \SetCell 的一列直接删除，并将 colspec 改为 {l c c c c}
& Module & Steps & Latency/step & Param. & FLOPs \\
\midrule
\SetCell[r=3]{c}
& PCCE & 1 & 1.38 ms& 22.5 K & 5.78 M  \\
& U-Net & 40 & 2.44  ms& 65.9 K & 9.55 M  \\%97.67
& FH & 20 & 0.27 ms& $-$ & 8.30 M \\%5.40
\midrule
\SetCell[r=2]{c} 
& JPDCE & 1 & 1.25 ms& 2.76 M & 6.69 M  \\
& SemS & 1 & $-$& 262 K & $-$  \\
\bottomrule
\end{tblr} 
\end{table}
Table~\ref{tab:model_complexity_tblr} shows the inference latency and complexity of different modules. 
All considered methods complete the pilot configuration within one time slot. The proposed method incurs an acceptable higher latency due to the diffusion process for more effective pilot configuration. Furthermore, the proposed SIW methods composed of the PCCE and conditional U-Net has only 22.5 K and 65.9 K trainable parameters, respectively, while FH introduces no additional learnable parameters, making the resulting predictor parameter-efficient.

\section{Conclusion} 
This paper proposed an SIW-based cognitive pilot design framework for semantic sensing in OFDM-ISAC systems. The SIW evaluates candidate pilot locations according to the residual uncertainty and task-dependent value of the corresponding future TF channel coefficients. Conditional diffusion with denoiser covariance estimation and local classifier sensitivity were used to estimate these two factors, respectively, enabling adaptive pilot allocation based on preceding observations. Experiments on CSI-Bench demonstrated consistent gains over static and adaptive pilot design baselines across different SNRs and pilot budgets. The results validate prospective semantic contribution estimation as an effective approach to resource-efficient semantic sensing in ISAC systems.

\section*{Appendix A: Bottleneck Approximation Derivation}
Denote $\mathbf g=\nabla_{\mathbf x}\mathcal L_{\rm task}(\mathbf x)$ and consider the Gaussian perturbation
$\boldsymbol\delta_{\boldsymbol\Gamma}
=\boldsymbol\Gamma^{-1/2}\boldsymbol\varepsilon$,
where $\boldsymbol\varepsilon\sim\mathcal N(\mathbf0,\mathbf I)$.
A second-order Taylor expansion gives
\begin{equation}
\mathbb E_{\boldsymbol\varepsilon}
\!\left[
\mathcal L_{\rm task}
(\mathbf x+\boldsymbol\delta_{\boldsymbol\Gamma})
\right]
\approx
\mathcal L_{\rm task}(\mathbf x)
+
\frac{1}{2}
\sum_{i=1}^{D}
\frac{[\mathbf H_{\rm task}]_{i,i}}{\gamma_i},
\label{eq:appendix_taylor_noise}
\end{equation}
where
$\mathbf H_{\rm task}=\nabla_{\mathbf x}^2
\mathcal L_{\rm task}(\mathbf x)$.
Using the gradient outer-product approximation
$[\mathbf H_{\rm task}]_{i,i}\approx g_i^2$ \cite{singh2020woodfisher}, the local IB objective for the $i$-th element becomes
\begin{equation}
J_i(\gamma_i)
\approx
\frac{g_i^2}{2\gamma_i}
+\beta\log\gamma_i,
\qquad
\gamma_i^\star = \frac{g_i^2}{2\beta}.
\label{eq:appendix_gamma_closed_form}
\end{equation}
Essentially, a channel element with a larger loss gradient indicates a higher task-relevant sensitivity, as perturbing the concerned element induces a larger local variation in the task loss. 
Since
$\left.\partial^2 J_i/\partial\gamma_i^2
\right|_{\gamma_i^\star}>0$,
the stationary point is a local minimum.
Restricting it to the diffusion-supported range yields
$\widehat\gamma_i^\star
=
\operatorname{clip}
\left(\frac{g_i^2}{2\beta},\gamma_{\min},\gamma_{\max}\right)$,
which gives \eqref{eq:gradient_gamma_star}.

 \bibliographystyle{IEEEtran}
\bibliography{refs_ieee_abbr.bib}

\end{document}